\documentclass[twocolumn,amsmath,amssymb,floatfix]{revtex4}

\usepackage{graphicx}
\usepackage{dcolumn}
\usepackage{bm}
\usepackage{subfig}
\usepackage{caption}
\begin{document}

\title{Emergence of Kinetic Behavior in Streaming Ultracold Neutral Plasmas}

\author{P. McQuillen, J. Castro, S. Bradshaw, and T.C. Killian}
\affiliation{%
Rice University, Department of Physics and Astronomy and Rice Quantum Institute, Houston, Texas 77005.}%
\date{\today}

\begin{abstract}
We create streaming ultracold neutral plasmas by tailoring the photoionizing laser beam that creates the plasma.   By varying the electron temperature, we control the relative velocity of the streaming populations, and, in conjunction with variation of the plasma density, this controls the ion collisionality of the colliding streams. Laser-induced fluorescence is used to map the spatially resolved density and velocity distribution function for the ions.
 We identify the lack of local thermal equilibrium and  distinct populations of interpenetrating, counter-streaming ions as
  signatures of  kinetic behavior. Experimental data is compared with results from a one-dimensional, two-fluid numerical simulation.
\end{abstract}


\maketitle

\section{Introduction}

The behavior and theoretical description of a plasma depends critically on the ratio of the mean free path for constituent particles  to the length scale of plasma inhomogeneities ($\left| n/\nabla n \right|$ for plasma density $n$, or $\left| T_{i,e}/\nabla T_{i,e} \right|$ for plasma ion or electron temperature $T_{i,e}$). For collisional systems in which  the mean free path is very small, particles remain close to local thermal equilibrium, and transport and collective motions can be described with a hydrodynamic treatment.
Because of the increasing mean free path with
particle kinetic energy, the situation can already change for mean free path  on
the order of or longer than 1\% of the length scale of plasma gradients \cite{gki80}. This gives rise to kinetic effects such as streaming plasmas and non-local transport, which are intensely studied because of their importance in many plasma environments, such as  high-density laser-produced plasmas \cite{pdd71,koo71,dms71,mob72,cgk73,gki80,lmp85,awb86,bcd88,ukm95,brt95,snb00,ggk04,bbr06,nfs06,rbb08,zrb09,mta13} and astrophysical plasmas like the solar wind \cite{pdf13}, solar atmosphere \cite{bra13,wbc08}, solar flares \cite{kde87}, and supernovae \cite{cga04}.

Here we show the emergence of kinetic behavior for ions in an ultracold neutral plasma \cite{kkb99,kpp07} by creating colliding plasma streams with controllable relative velocity and density.
It is the first observation in ultracold neutral plasmas of ion transport \cite{kkb00,lgs07,cmk10,kmo12,mcs13} that can not be explained with a hydrodynamic treatment.
The emergence of kinetic effects is heralded by the absence of local thermal equilibrium for ions and interpenetrating streaming plasma populations.  This compliments recent work on acoustic-wave phenomena in ultracold neutral plasmas in the purely hydrodynamic regime \cite{cmk10,kmo12,mcs13}. Results are compared with a two-fluid hydrodynamic code \cite{bma03,bkl11}.

\begin{figure*}[ht]
\includegraphics[width=7in]{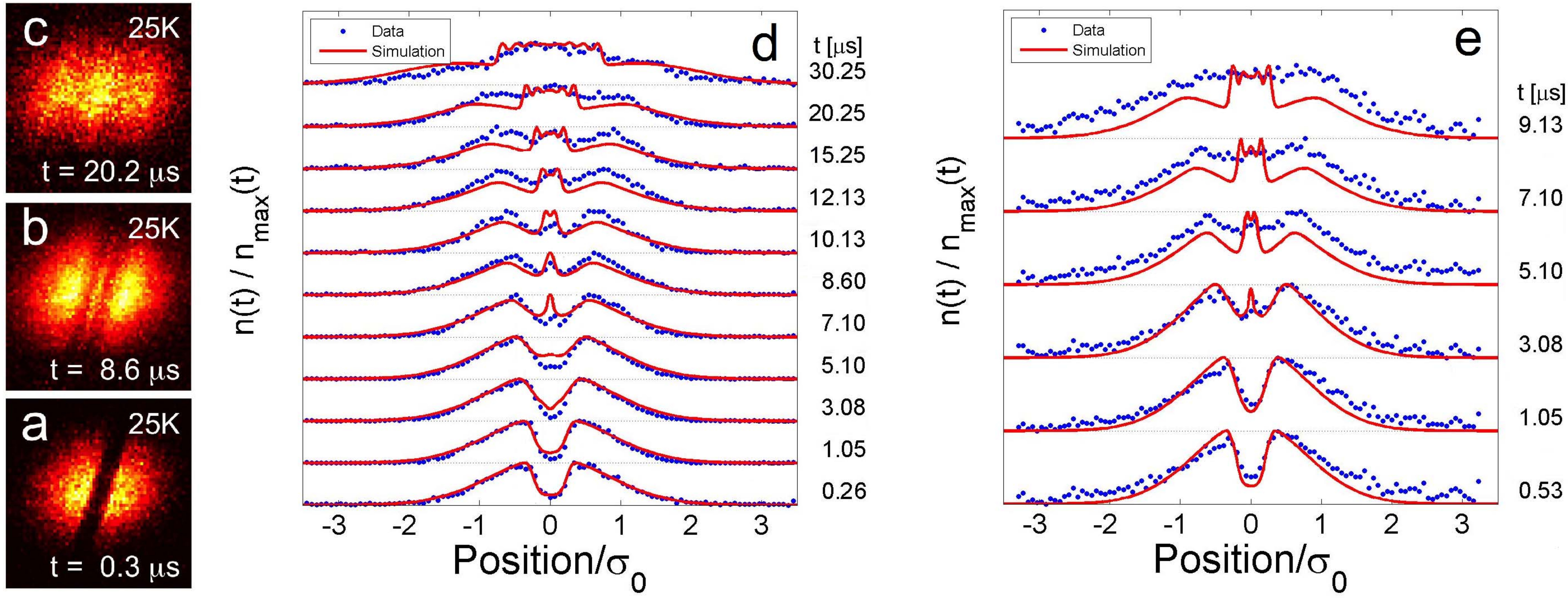}
\caption{
 Evolution of the ion density distributions.
 The time since ionization is indicated. Initial peak densities are $6\times 10^{14}$\,m$^{-3}$.
  (a-c)  False color plots of the density distribution in the crossover between hydrodynamic and kinetic regimes ($T_e(0)=25$\,K).
  (d) 1D slices through the density profile for the same data as (a-c), normalized and offset for clarity.  Note the emergence of the central feature as the gap splits. The hydrodynamic simulation overestimates the prominence of the central density enhancement and the density depletions propagating away from the gap region.
  (e) Same as (d)  but for $T_e=105$\,K  showing more kinetic behavior. Note the absence of a significant central feature at the confluence of the ions streaming into the gap and the poor agreement between the data and hydrodynamic simulation.
  Distance is normalized to plasma size $\sigma_0=1.5\,$mm.
  }
\label{fig:ImagesandProfiles}
\end{figure*}

\section{Experimental setup}
\label{sec:methods}

Ultracold neutral plasmas are created through photoionization of strontium atoms from a magneto optical trap (MOT) as described in \cite{scg04}.
The initial electron temperature  is tunable ($T_e(0)\approx1-1000$\,K), and is determined by the excess energy of the ionizing photons above threshold. Ion temperatures are set to approximately 1\,K by disorder-induced heating \cite{mur01,csl04}, resulting in strongly-coupled ions in the liquid-like regime \cite{kpp07,ich04}.  The plasma density distribution is a spherical Gaussian, $n(r)=n_0\mathrm{exp}(-r^2/2\sigma_0^2)$, with  $n_0\sim$\,$6\times10^{14}$\,m$^{-3}$ and $\sigma_0\sim$\,1.5\,mm.

A gap in the center of the plasma is created by placing an opaque wire in the path of the ionizing beam, which results in
 two plasma hemispheres that stream into each other during plasma expansion.
 Streaming plasmas are precursors to complex phenomena such as instabilities, shock waves, and solitons \cite{mel86,cap94,bcd88,rbb08}, and they have been studied in a wide variety of environments
such as the solar wind \cite{gom87},
supernovae \cite{cga04}, 
 double plasma devices \cite{gdo75,nni77,nbi04,sss05},
and laser-produced plasmas \cite{pdd71,mob72,bcd88,rbb08,koo71,dms71, cgk73}.
A partially transmitting strip, instead of an opaque wire, was recently used to create a smaller amplitude density depletion in the plasma center \cite{mcs13}. This led to hydrodynamic phenomena such as gap splitting and propagation of localized density depletions at the ion acoustic wave speed. A similar technique was  used to create periodic density modulations and excite ion acoustic waves \cite{cmk10}.

As a diagnostic, a tuneable, narrowband  laser induces fluorescence from the primary Sr$^+$ transition, $^{2}$S$_{1/2}-^{1}$P$_{1/2}$ at $\lambda=$\,422\,nm \cite{lgs07}. This excitation beam is masked by a 1\,mm slit to illuminate only a central sheet of the plasma \cite{cgk08}. Fluorescence emitted close to perpendicular to the plane of the ionizing and 422\,nm beams is imaged onto an intensified CCD camera with a pixel size of $52\,\mu$m.
The resulting image, $F(x,y,\nu)$, is frequency dependent due to the natural linewidth and Doppler-broadening of the transition \cite{cgk08}.
Images are combined to obtain a map of
the density distribution of the ions in the illuminated plane, $n_i(x,y,z\approx 0)$
or the spatially resolved,  distribution function for  velocity perpendicular to the gap \cite{cgk08,cmk10}.

We also model the plasma evolution with a one-dimensional two-fluid simulation \cite{bma03,bkl11} in which we treat bulk flows (transport, compression, and rarefaction), energy exchange between particle species by collisions and the work done by  small-scale electric fields. We found that thermal conduction and viscous interactions made negligible contributions to the plasma energy balance in the temperature regimes explored here. The initial density distribution is matched to the profile observed experimentally within 100\,ns of plasma creation. Ion temperature is set equal to the value  after disorder-induced heating \cite{csl04} measured with LIF spectroscopy \cite{cgk08}. A one-dimensional simulation will not correctly reproduce the adiabatic cooling of electrons and ions caused by the three dimensional expansion of the ultracold plasma \cite{lgs07}. But the deviation is small and does not affect the phenomena discussed here.

The critical parameter that demarcates  hydrodynamic (collisional) and kinetic (collisionless) regimes is the ratio between the mean free paths of the particles and a characteristic scale-length of the plasma (the Knudsen number). When the mean free path is small compared with this scale length then the plasma is collisional and the particle velocity distribution functions are near Maxwellian. When the mean-free-path exceeds the characteristic scale-length then kinetic phenomena can become important.

 Mean free paths and collision timescales for ion-ion and electron-electron collisions respectively for conditions relevant to strontium ultracold plasmas  are defined in terms of the generalized collision frequency for a test particle ($\alpha$) streaming with speed $v_{\alpha}$ through a background of field particles ($\beta$)  with and  mass $m_{\alpha,\beta}$:
\begin{equation}
\nu^{\alpha,\beta} = \left(1 + m_{\alpha} / m_{\beta} \right) \psi\left( x^{\alpha,\beta} \right) \nu_0^{\alpha,\beta},
\label{eqnSB10}
\end{equation}
\noindent where
\begin{equation}
\psi(x) = \frac{2}{\sqrt{\pi}} \int_0^x t^{1/2} \exp(-t) dt,
\label{eqnSB11}
\end{equation}
\begin{equation}
x^{\alpha,\beta} = \frac{m_{\beta} v_{\alpha}^2}{2 k_B T_{\beta}},
\label{eqnSB12}
\end{equation}
\noindent and
\begin{equation}
\nu_0^{\alpha,\beta} = \frac{ e_{\alpha}^2 e_{\beta}^2 \lambda_{\alpha,\beta} n_{\beta}}{4\pi \epsilon_0^2 m_{\alpha}^2 v_{\alpha}^3}.
\label{eqnSB13}
\end{equation}
$\lambda_{\alpha,\beta}$ is the Coulomb logarithm for the interaction.
In the absence of a gap, at the density of the unperturbed plasma,  the ion-ion mean free path for ions with four times the thermal speed for $T_i=1$\,K  is 40\,$\mu$m for screening by  100\,K electrons. This is small compared to the plasma size $\sigma_0$, and we expect predominantly hydrodynamic behavior for the ion motion such as overall plasma expansion. Electron mean free paths are much longer, which might become important in higher temperature plasmas where the streaming population could influence the energy balance and, consequently, the dynamics of the plasma via high order moments of the distribution function, such as the heat flux. However, for the parameter space and phenomena explored here, we are not sensitive to these effects.


\section{Results and Discussion\label{sec:RnD}}

A qualitative understanding of the plasma dynamics when a gap is formed can be gained from looking at the evolution of the ion density distributions. Figure \ref{fig:ImagesandProfiles} shows data for conditions that display phenomena characteristic of the crossover regime from hydrodynamic to kinetic ($T_e(0)=25$\,K) and the fully kinetic regime ($T_e(0)=105$\,K). The initial density distributions are similar for both cases.
Figures \ref{fig:ImagesandProfiles}(a)-(c) shows false color images of the two-dimensional central density slice $n_i(x,y,z\approx 0)$, and
Figs.\  \ref{fig:ImagesandProfiles}(d) and (e) show 1D central density traces at various times for initial electron temperature (d) $T_e(0)=25$\,K and (e) $T_e(0)=105$\,K.
(Due to optical access constraints, the plasma is slightly larger than the region illuminated by the fluorescence beam, creating the apparent oblong shape in the 2D images.)

Immediately after plasma formation, a gap is evident (Figs.\ \ref{fig:ImagesandProfiles}(a), (d), and (e)) with
a length scale for the sides of the gap of about $l \equiv\left| n/\nabla n \right|\sim$\, 400\,$\mu$m, which is determined by diffraction of the ionizing beam and is much less than $\sigma_0$.
The subsequent movement of ions into the gap 
 can be understood hydrodynamically using the coupled ion-electron momentum balance equations \cite{kkb00}. In the presence of any density gradient, ions are accelerated ($a_0$)  by the electron pressure ($P_e$) according to $Mna_0\approx-\nabla P_e\approx-\nabla\left(nk_BT_e(0)\right)$. $T_e(0)$ is expected to be close to uniform \cite{rha03}, and a characteristic ion velocity after acceleration is $v_f\approx\sqrt{k_BT_e(0)/M}$. This suggests a timescale for motion of
\begin{equation}
\tau\approx \frac{v_f}{a_0}\approx -\frac{n}{\nabla n}\sqrt{\frac{M}{k_BT_e(0)}}.
\end{equation}
At the sides of the gap, this timescale is fast and can be identified with the timescale for ion streams to form and interact in the center of the gap, $\tau_{fast}\approx\sqrt{Ml^2/[k_BT_e(0)]}$. Here,   $\tau_{fast}\sim$\,8\,$\mu$s for $T_e(0)$=25\,K and $\tau_{fast}\sim$\,4$\mu$s for $T_e(0)$=105\,K, which approximately matches the timescales seen in the experiment.
Note that this argument also explains the overall expansion of the plasma, for which the length scale is that of the spherical gaussian, $-n/\nabla n\approx\sigma_0$. This gives the characteristic expansion time of a self-similar expansion, $\tau_{exp}\approx\sqrt{M\sigma_0^2/[k_BT_e(0)]}$ \cite{kpp07}
($\tau_{exp}= 15$\,$\mu$s and 31\,$\mu$s for $T_e(0)$=25\,K and $T_e(0)$=105\,K respectively).

\subsection{Signatures of the Kinetic Regime in the Density Evolution \label{sec:DensityEvolution}}

Whether the ion streams behave hydrodynamically or kinetically when they collide depends upon the inter-stream mean free path at the confluence. The relevant density-gradient length scale for this interaction  can be approximated as something between the  length scale of the side of the gap ($l$) and the gap width. For short mean free path, the streams cannot penetrate, and a density enhancement emerges at the confluence. The gap splits and two density depletions  propagate away from plasma center as seen in  \cite{mcs13}. There is indication of this behavior for $T_e(0)=25$\,K (Figs.\ \ref{fig:ImagesandProfiles}(b-c)). However, the central feature never rises to the height that the plasma would have had in the absence of an initial gap, and the amplitude of the features moving away from the center are much smaller than the initial gap (Fig.\ \ref{fig:ImagesandProfiles}(d)). These two caveats indicate that the colliding streams in the $T_e(0)=25$\,K  plasma are not fully in the hydrodynamic regime and kinetic effects are important. Experiments in \cite{mcs13} accessed and characterized the purely hydrodynamic regime for $T_e(0)=25$\,K by starting with a very small density depletion and approximately five times greater density in the gap. In this case, the central
feature rose to full height, and the amplitude of the traveling density depletions showed no significant attenuation.

To estimate the mean free path $\ell_{mfp}$ for an ion in one stream moving through the other, we use the density in the center at time $\sim \tau_{fast}/2\sim$\,4\,$\mu$s, which is about $10^{14}$\,m$^{-3}$, and take the velocity as $v_f\approx\,50$\,m/s. Equation \ref{eqnSB13} yields $\ell_{mfp}=600\,\mu$m.

To highlight the difference between the density evolution in the experiments and what one would predict for hydrodynamic behavior, we show in Fig.\ \ref{fig:ImagesandProfiles}(d) the results of a 1-D hydrodynamic simulation of the experimental data. For the hydrodynamic plasmas studied in \cite{mcs13}, the  simulation accurately captured all the features seen in the experiment. Here, there is obvious discrepancy between experiment and the hydrodynamic simulation in the gap region. The simulation overestimates the prominence of the central feature and traveling depletions.

The mean free path is further increased by increasing the electron temperature and relative velocity of the colliding streams, as shown in Fig.\ \ref{fig:ImagesandProfiles}(e) for $T_e(0)$=105\,K. Making the same approximations as before yields $\ell_{mfp}=7$\,mm.
In this case, the hydrodynamic phenomena related to the gap (central
density enhancement and traveling density depletions) are barely visible, indicating the dominance of kinetic effects.
There is also very large discrepancy between experiment and the hydrodynamic simulation in the gap region.


\subsection{Measuring the Velocity Distribution Function \label{sec:VelDistribution}}

To probe the effect of the gap on the plasma dynamics more quantitatively, and to support our statements regarding the emergence of kinetic behavior, we analyze the fluorescence excitation spectrum.
We obtain the spectrum of a region of the plasma, $ S(\nu)_{region}$, by summing the fluorescence from that region,
\begin{equation}
\label{eq:Fluorspec}
 \sum_{x,y\in region} F(x,y,\nu)\propto S(\nu)_{region}.
\end{equation}
We will first discuss the spectrum of the entire illuminated plasma sheet, which is obtained by expanding the region to encompass the entire field of view. Then  we will discuss the spectra of resolved regions of the plasma near the gap.

\begin{figure}[ht]
\begin{center}
\includegraphics[width=3.4in]{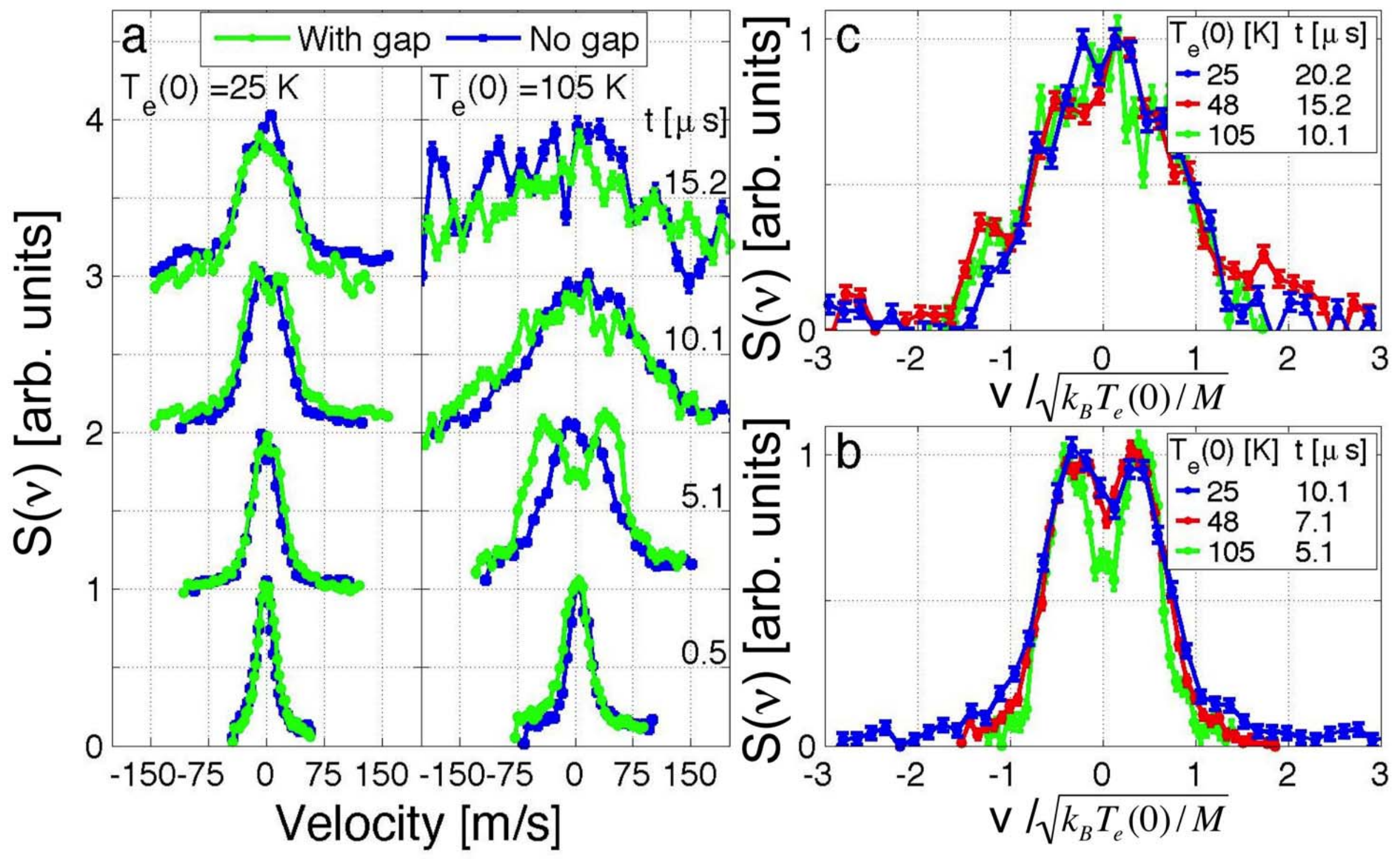}
\caption{Spectra of entire plasma sheet. (a) Effect of the gap and changing $T_e(0)$. Frequency has been converted to velocity ($\mathrm{v}$) through the Doppler shift $\mathrm{v}=\nu/\lambda$, and time since ionization is indicated.   Note the emergence of bimodal structure, corresponding to streaming plasmas. 
(b) Spectra of plasmas with gaps  with scaled velocity and  $t/\tau_{fast}\sim1.3$. The universal behavior highlights the emergence of the new timescale, $\tau_{fast}$.
(c) Same as (b) but for $t/\tau_{exp}\sim 0.7$. At later times, structure has vanished, and universal behavior indicates the scaling of the transfer of electron energy into plasma expansion velocity.}
\label{fig:WholeCloudSpec}
\end{center}
\end{figure}

The spectral shape is dominated by Doppler shifts and broadenings resulting from ion velocities, and for simplicity we can neglect the  small broadening of the spectra due to the natural linewidth of the transition. This allows us to interpret the spectrum
 as the distribution of ion velocities perpendicular to the gap by plotting the spectra versus velocity, where velocity is calculated as $\rm{v}=\delta \nu \lambda$. $\delta \nu$ is the detuning of the fluorescence excitation laser from resonance and $\lambda$ is the wavelength. (The FWHM,  natural linewidth of the transition $\gamma/2\pi=20$\,MHz leads to a width
 in velocity of $\delta \mathrm{v}=\gamma\lambda/2\pi=8.4$\,m/s.)

\subsection{Velocity Distribution of the Entire Plasma \label{sec:DensityEvolution}}

Using this approximation, Fig.\ \ref{fig:WholeCloudSpec}a shows the velocity distribution for the entire plasma sheet, which reflects directed ion motion (plasma expansion or streaming) and thermal ion motion.  At early times, the distribution is unaffected by the existence of the gap, and the width reflects the ion temperature after disorder-induced heating (DIH) \cite{mur01,csl04,cdd05}.
DIH occurs as excess potential energy in the initially random ion spatial distribution is converted to ion kinetic energy. The timescale for DIH is the inverse of the ion plasma oscillation frequency, $1/\omega_{pi} =\sqrt{M\epsilon_0/ne^2}\sim\,1\,\mu$s, where $M$ is the ion mass. This is much faster than the hydrodynamic timescale on which the plasma reacts to density inhomogeneities.
After DIH, ion velocities increase as electron thermal energy is converted to directed ion motion through the hydrodynamic process discussed above.  The  distribution for a plasma without a gap broadens as the ions accelerate radially, but it retains a generally featureless Gaussian shape characteristic of a self-similar expansion \cite{lgs07}.

The spectrum of the perturbed plasma broadens more quickly and develops a bimodal shape. We will show below that bimodality reflects ions from the two sides of the gap streaming into the gap and through each other. This motion is still driven by the conversion of electron thermal energy into ion expansion energy, but it occurs on the timescale $\tau_{fast}$. Figure \ref{fig:WholeCloudSpec}b shows spectra for the perturbed plasma at the same scaled time, $t/\tau_{fast}\sim1.3$ and with velocity scaled by $v_f$. The universality of the scaled data emphasizes how the gradient in the electron thermal pressure drives the dynamics.
Note that at later times 
(Fig. \ref{fig:WholeCloudSpec}a top panel) the distributions for the perturbed and unperturbed plasmas display the same scaled width with no sign of bimodality.
The structure in the global velocity distribution has vanished even though some structure in the density distribution still persists (Fig. \ref{fig:ImagesandProfiles}).

\subsection{Spatially Resolved Velocity Distributions and Streaming Plasmas \label{sec:DensityEvolution}}

\begin{figure}[ht]
\begin{center}
\includegraphics[width=3.5in]{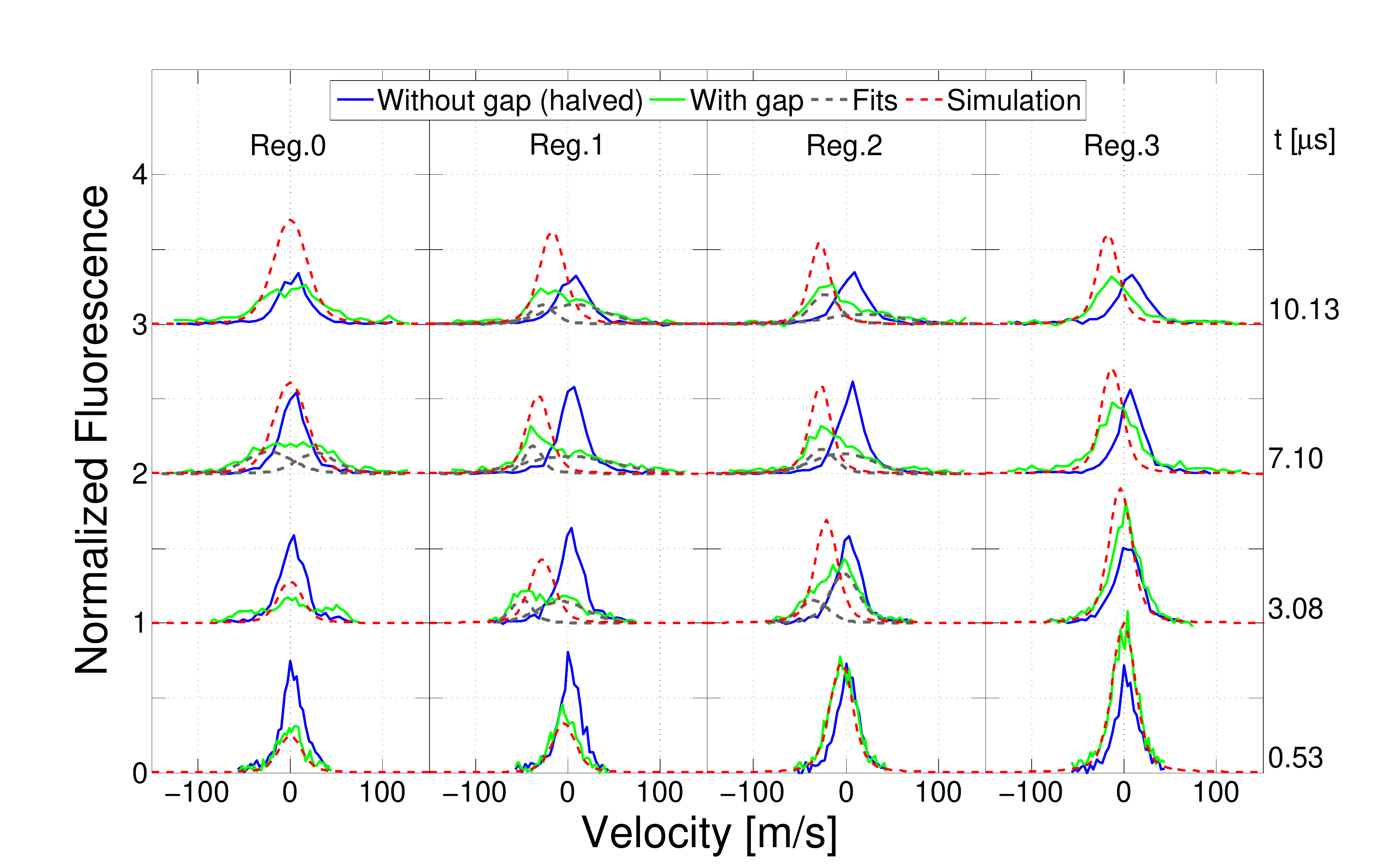}
\caption{Spatially-resolved regional spectra for  plasmas with $T_e(0)=25$\,K. Time since ionization is indicated on the right. The solid lines are data with and without the gap while short dashed lines are fits to perturbed velocity distributions. Shifts represent ion mean velocities and widths represent ion temperature. For plasmas without the gap, positive velocities reflect expansion \cite{cgk08}. For plasmas with the gap, multiple populations appear moving left and right.  Time since ionization is indicated on the right.  The long dashed lines show results of the hydrodynamic simulation.
}
\label{fig:RegionSpectra25}
\end{center}
\end{figure}

\begin{figure}[ht]
\begin{center}
\includegraphics[width=3.5in]{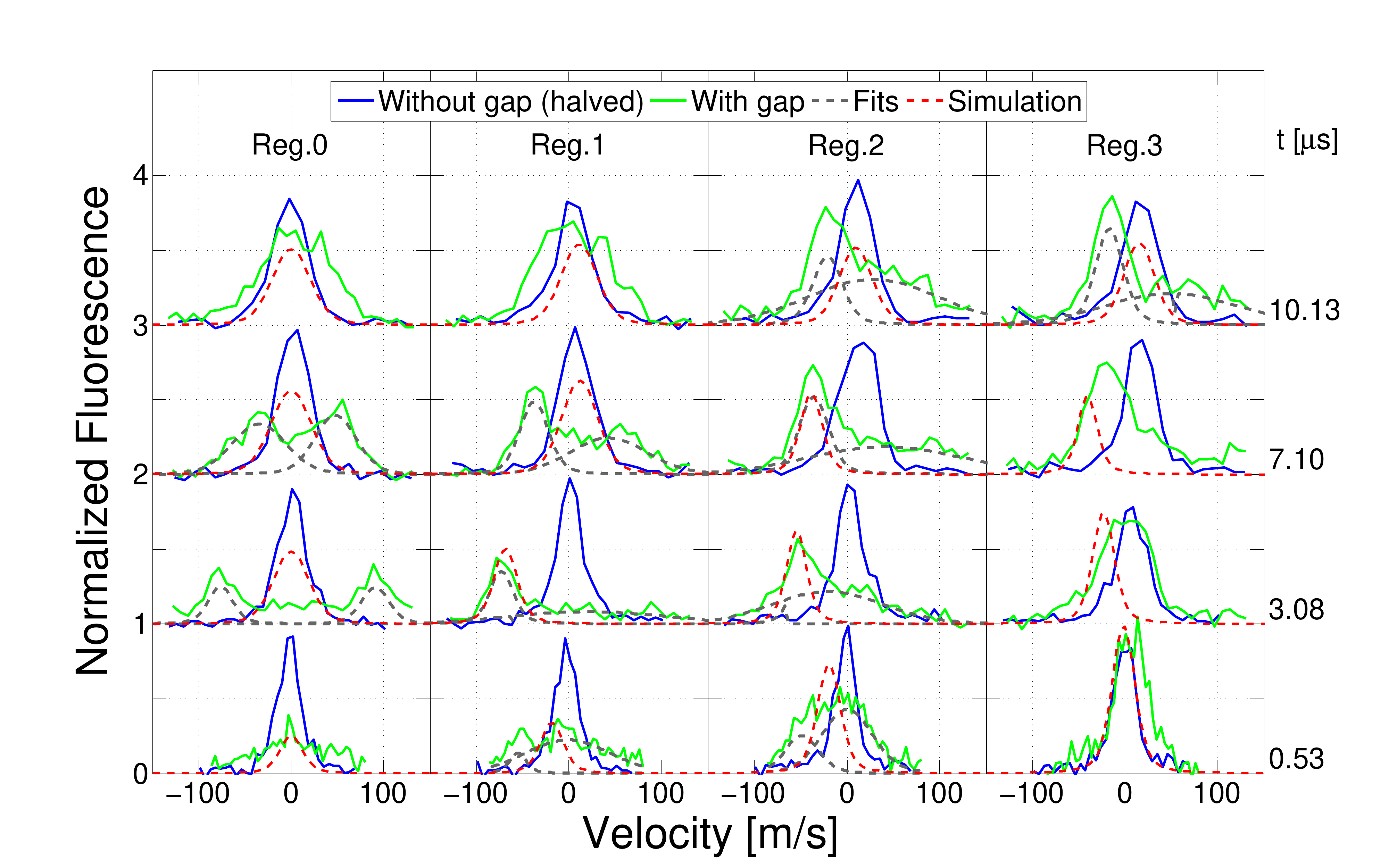}
\caption{Same as Fig.\ \ref{fig:RegionSpectra25}, but for $T_e(0)=105$\,K. Compared to Fig. \ref{fig:RegionSpectra25}, in the perturbed plasma, average velocities are much higher and populations moving to the right are more prominent.
}
\label{fig:RegionSpectra105}
\end{center}
\end{figure}

\begin{figure}[ht]
\begin{center}
\includegraphics[width=3.2in]{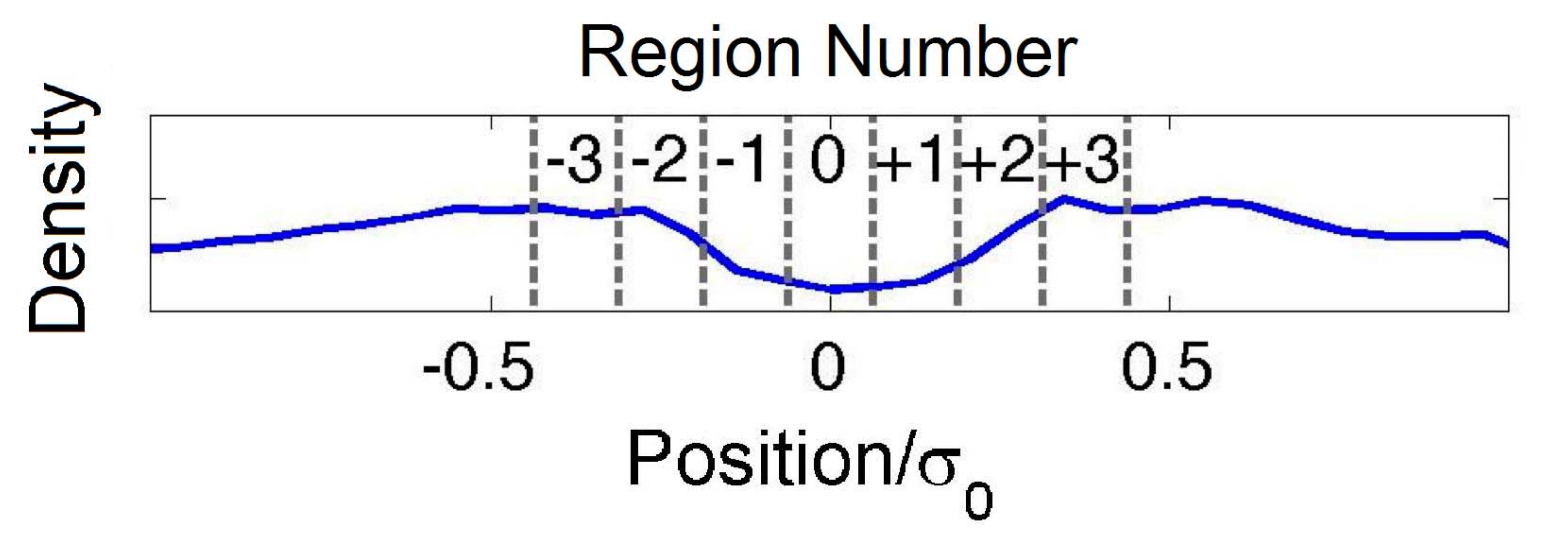}
\caption{Regions used in analysis of local ion velocities and temperatures.}
\label{fig:Regions}
\end{center}
\end{figure}

To probe the local dynamics of the plasma, we analyze spectra and extract the 1D velocity distributions of smaller spatial regions  in the vicinity of the gap, which measure local ion velocities and temperatures \cite{gls07,lgs07,cgk08} (Figs. \ref{fig:RegionSpectra25} and   \ref{fig:RegionSpectra105} for initial electron temperatures 25\,K and 105\,K respectively).
Due to the symmetry of our plasma with respect to the gap, corresponding regional spectra in the two hemispheres mirror each other. We only show the spectra of the central region and the regions to the right of the gap. Spectra are obtained from Eq. \ref{eq:Fluorspec}, where $region$ refers to the area covered by regions 0 to 3 in Fig. \ref{fig:Regions}.  The dimensions of these regions are 190\,$\mu$m perpendicular to the gap and 760\,$\mu$m parallel to it.
 The shift of a distribution represents the mean velocity of an ion population, and the width represents the rms velocity spread \cite{cgk08}. We typically quote the width in terms of an effective local temperature, but the width is also affected by spread in average ion velocity across the region and lack of thermal equilibrium.

Without the gap, regional spectra show expansion velocities that increase with time and distance from center  as expected \cite{kpp07}.
 For regions far from the gap and especially at early times, the distribution shape, shift, and width are essentially unchanged by introduction of a gap.
 But there are dramatic differences between spectra of perturbed and unperturbed plasmas for inner regions.

 Signal at positive velocities represent ions moving away from plasma center, whereas, negative velocities represent ions moving towards it. Positive and negative velocity groups in region 0  indicate ions moving through the center of the gap. It is important to note that ions move through the regions, so spectra for a given region at different times may reflect different ions. For instance, neglecting collisions and given the dimension of the regions, particles in a particular region with mean velocity of 60\,m/s, will be in a contiguous region in approximately $\sim3\,\mu$s.

There are many common features in $T_e(0)=25$\,K and $T_e(0)=105$\,K data that illustrate the general  behavior of streaming ultracold neutral plasmas.
At early times ($0-3\,\mu$s), the density gradient in regions 1 and 2 produces an acceleration of ions  towards the gap. This is followed by the appearance of two beams of high velocity ions crossing each other in region 0.
The stream velocities closely match the characteristic velocity  introduced earlier ($v_f=\sqrt{k_B T_e(0)/M}=100$\,m/s and 50\,m/s for 105\,K and 25\,K respectively). The bulk velocity of each population greatly exceeds the ion thermal speed and is comparable to the  phase velocity of ion acoustic waves.
At later times ($7-10\,\mu$s) these populations have crossed region 0, and ions  that originated on the other side of the gap now appear in regions 1-3 as ions moving away from the gap. These interpenetrating streams and local velocity distributions that cannot be defined by a single Maxwellian with a well-defined temperature are indicators that the plasma is in the kinetic regime. Data from hydrodynamic plasmas studied in \cite{mcs13} showed no deviation from local Maxwellians.

\begin{figure}[ht]
\begin{center}
\includegraphics[width=3.5in]{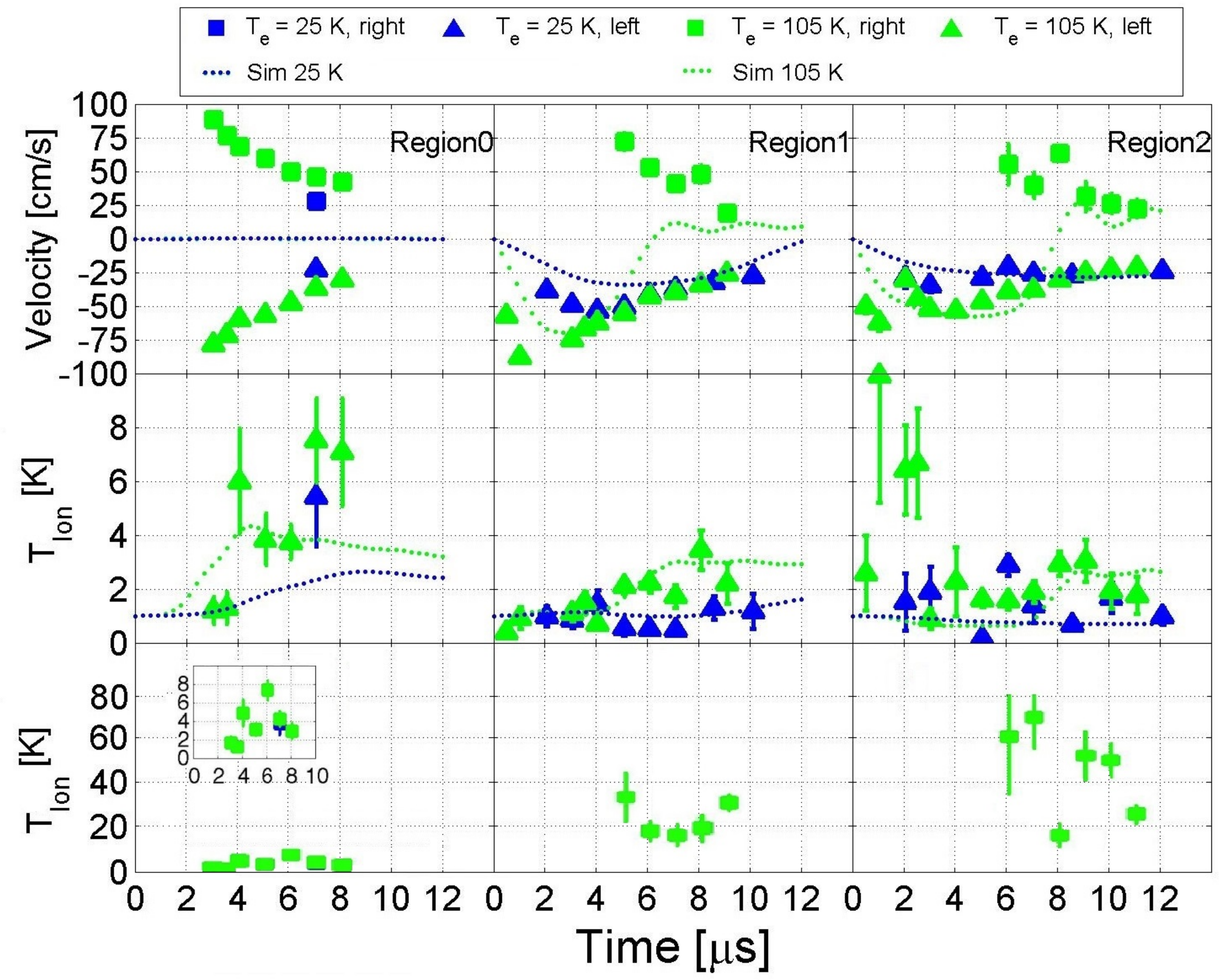}
\caption{ Ion velocities and temperatures for the fast moving streams in Figs. \ref{fig:RegionSpectra105} and \ref{fig:RegionSpectra25}. (top) Velocities of the fast moving streams. Note that streams moving towards the gap (velocity $<0$) appear first in region 2. Streams moving away from the gap velocity $>0$) appear first in region 0. (middle) Temperature of left-moving streams moving towards the gap. (bottom)  Temperature of right-moving streams that have crossed the gap and are moving away from it. Note the different temperature scales. }
\label{fig:VelAndTemperture}
\end{center}
\end{figure}

In Figs.\ \ref{fig:RegionSpectra25} and  Figs. \ref{fig:RegionSpectra105}, it is interesting to note that ion populations remain cold 
until they have passed through the gap, after which their velocity spread greatly increases (Fig. \ref{fig:VelAndTemperture}).
Ions remaining in the gap or moving in at later times (10\,$\mu$s) eventually merge into one population of hot ions.

Important differences between the data sets reflect the variation in electron thermal pressure. For example, moving populations appear sooner for $T_e(0)=105$\,K ($0.5\,\mu$s) than $T_e(0)=25$\,K, and velocities are higher for higher $T_e(0)$ (Fig. \ref{fig:VelAndTemperture}). The ion temperature in region zero at later times is also hotter for $T_e(0)=105$\,K than $T_e(0)=25$\,K.

There is also significant difference in the penetration depth of ions from one side into the plasma on the other side of the gap. We estimate the penetration length in the plasma from the farthest excursion of right-moving populations. For $T_e(0)=105$\,K, right-moving ions clearly appear in region 3 (10\,$\mu$s), while for $T_e(0)=25$\,K right-moving ions appear in region 1, and perhaps region 2, and the population is less distinct than for $T_e(0)=105$\,K. $T_e(0)=48$\,K (not shown) displays intermediate behavior.
No clear right-moving population can be found at times later than $\sim$10\,$\mu$s for any $T_e(0)$ (data not shown). One expects a greater penetration length for higher $T_e(0)$ because the Coulomb collision cross section decreases with  collision velocity, $\sigma \propto 1/v^4$. From these measurements, it should also be possible to estimate the stopping power of electrons and strongly coupled ions in an ultracold plasma for a penetrating ion stream \cite{pme91,bps05,tbd08,ggm08}.


Hydrodynamic simulations are included in Figs.\ \ref{fig:RegionSpectra25} and  Figs. \ref{fig:RegionSpectra105} to emphasize the breakdown of the hydrodynamic approximation. But it is also interesting to note  the nature of the discrepancy between the data and the simulation. In the central region, when two symmetric, relatively cold, streaming plasma populations are present, the hydrodynamic code predicts a single cold population with zero average velocity, which is what one would expect given the constraint in the simulation of a local thermal equilibrium for the  ions. However, in regions away from the center,  the distributions from the hydrodynamic simulation are close the the distribution for the population of ions that have not crossed the gap, but the contributions from ions that have streamed across the gap are absent in the hydrodynamic simulation. This reflects the inability of the hydrodynamic code to account for non-local transport. The hydrodynamic code also tends to underestimate the ion temperature at later times. 


\subsection{Trends in the Ion Temperature and Bulk Velocity} \label{sec:TempandVelocity}

To accurately extract ion temperatures and velocities we fit regional spectra (Figs. \ref{fig:RegionSpectra105} and \ref{fig:RegionSpectra25}) to  one or the sum of two Voigt profiles.
The Voigt profile is the convolution of a Lorentzian with homogeneous width  $\gamma$ and shifted Gaussian for Doppler effects \cite{kpp07}. In this way we take spectral broadening due to the natural linewidth of the transition into account for quantitative analysis. As mentioned above, the temperature is extracted from the width of the Gaussian. Temperature measurements are sensitive to many systematic effects,
but it still an informative parameter.
Data that fit well to two profiles give clear indication of different populations of streaming ions with different mean velocities and temperatures. In these cases, the dotted lines show the individual Voigt profiles of each population, and the sum of both these curves is in good agreement with the data.

Ion velocities and temperatures for the  streams are shown in Fig. \ref{fig:VelAndTemperture} whenever they are resolvable.
For higher $T_e(0)$ we obtain faster streaming velocities, which is consistent with the separation of the bimodal distribution in Fig. \ref{fig:WholeCloudSpec}a.  In regions 1 and 2, populations that are moving into the gap (negative velocities and middle panel) appear to be relatively cold ($T_i\lesssim 2$\,K), while populations that have traveled across the gap  and moved an appreciable distance through a counter-streaming plasma (positive velocities and bottom panel), have higher temperatures ($T_i\gtrsim 20$\,K). In the center, region 0, particles moving to the left and right show very similar ion temperatures.

For all regions, velocities decrease with time. Decreasing plasma gradient and decreasing $T_e$ \cite{lgs07} with time imply that ions moving through the gap at later times have experienced less acceleration during their trajectories than at early times. But velocities also decrease for an individual population as it moves through the regions.  This may reflect that after ions move past the gap, the local electric field opposes the average motion for that population.  Damping or collisions may also play a role, which is supported by the considerably higher temperatures measured for populations that have crossed the gap.

When only one distinct population of ions is visible in the velocity distribution function, such as in region 1 at early times ($<3\,\mu$s), the hydrodynamic simulation does a reasonable job of predicting the average ion velocity (\ref{fig:VelAndTemperture}(top)). When two streaming populations are present, however, such as in region 1 at later times or in region 0, the velocity distribution is far outside the regime that can be treated by a hydrodynamic treatment. The simulation tends to track close to the average velocity, which is always zero in region 0. Similarly, in regions farther from the gap (region 2), the temperature in the simulation agrees reasonably well with the temperature of the population moving towards the gap (Fig.\ \ref{fig:VelAndTemperture}(middle)), indicating these ions are relatively unaffected by the small population of high velocity ions that have crossed the gap and are streaming through the region. This highlights the phenomena that characterize departure from the hydrodynamic regime.



\section{Conclusion \label{sec:Conclusion}}
In this work we have used the ability to create streaming populations of particles in ultracold neutral plasmas to demonstrate the crossover from hydrodynamic to kinetic behavior and predominantly kinetic behavior. This work is complimentary to a recent paper in which we studied plasmas with a similar initial geometry in the hydrodynamic regime \cite{mcs13}. The ratio of the ion mean free path to the length scale of plasma inhomogeneities   as ions move through the gap and the populations collide is the crucial parameter for determining the regime of behavior, and it can be tuned by varying the ion density in the gap and the electron temperature, which changes the stream velocity. The hydrodynamic regime is characterized by a density enhancement at the confluence of the plasma streams, gap splitting, and localized density depletions that propagate away from the initial gap. The kinetic regime lacks these features and displays velocity distributions that strongly deviate from local thermal equilibrium and show distinct populations of interpenetrating, counter-streaming plasmas.

There are many future directions in which this work can go. This study has focused on a qualitative description of plasma behavior in the various regimes, but it would be informative to  quantitatively map the boundary between different regimes in the parameter space of stream velocity and plasma density in the gap. One could use, for example, the appearance of streaming populations or visibility of a localized density enhancement in the confluence region as an indicator of the plasma regime.

A full kinetic description will allow a much more detailed and quantitative discussion of the observed effects. Alternatively, data such as this can provide a stringent test of kinetic codes. Kinetic codes, however, are much more computationally intensive.
Rather than move to a full kinetic description, a common strategy is to generalize transport coefficients in order to extend the usefulness of hydrodynamic treatments.  This is often done to treat
nonlocal transport in plasmas with low collisionality, in which particle and energy fluxes at a given location are affected by conditions in distant regions of the plasma, which represents an important and complex problem.
Often, \textit{ad hoc} schemes are used in which the flux  is simply limited to a fraction  of its free-streaming value \cite{mld94,ggk04,ylw10,gki80,awb86,lar93}. Data such as presented here can provide a measurement of fluxes and thermodynamic gradients with high temporal and spatial resolution. This could provide a  test of flux-limiting treatments or inform improved schemes to extend hydrodynamic models to address nonlocal effects.

We have mentioned that it should be possible to estimate the stopping power of ions by studying the penetration depth of the ion streams. Away from the initial gap, the streams move into regions of cold ions that are strongly coupled, which gives greater interest to such a measurement \cite{pme91,bps05,tbd08,ggm08}.
With sharper density features in the initial plasma, it may also be possible to excite and study shock waves \cite{hua02,rha03}.

This work was supported by the Department of Energy and National Science Foundation (PHY-0714603). SJB acknowledges funding support from NASA.


\end{document}